\documentclass[12pt,preprint]{aastex}

\shortauthors{Rhee et al.}

\begin{document}

\title{EF~Cha: Warm Dust Orbiting a Nearby 10 Myr Old Star}

\author{Joseph H. Rhee\altaffilmark{1}, Inseok Song\altaffilmark{2}, 
B. Zuckerman\altaffilmark{3}}

\altaffiltext{1}{Gemini Observatory, 670 North A'ohoku Place, Hilo, HI 
96720; current address (Department of Physics and Astronomy, Box 951547,
University of California, Los Angeles, CA 90095-1562; rhee@astro.ucla.edu)}
\altaffiltext{2}{Gemini Observatory, 670 North A'ohoku Place, Hilo, HI 
96720}
\altaffiltext{3}{Department of Physics and Astronomy, Box 951547,
University of California, Los Angeles, CA 90095-1562; NASA Astrobiology 
Institute}

\begin{abstract}
Most Vega-like stars have far-infrared excess (60\,$\micron$ or 
longward in {\it IRAS, ISO, or Spitzer MIPS} bands) and contain cold 
dust ($\lesssim$150\,K) analogous to the Sun's Kuiper-Belt region.  
However, dust in a region more akin to our asteroid belt and thus relevant 
to the terrestrial planet building process is warm and produces excess 
emission in mid-infrared wavelengths.  By cross-correlating Hipparcos 
dwarfs with the MSX catalog, we found that EF~Cha, a member of the 
recently identified, $\sim$10\,Myr old, ``Cha-Near'' Moving Group, 
possesses prominent mid-infrared excess.  N-band spectroscopy reveals 
a strong emission feature characterized by a mixture of small, warm, 
amorphous and possibly crystalline silicate grains.  Survival time of 
warm dust grains around this A9 star is $\lesssim 10^{5}$ yrs, much 
less than the age of the star.  Thus, grains in this extra-solar terrestrial 
planetary zone must be of "second generation" and not a remnant of 
primodial dust and are suggestive of substantial planet formation 
activity.  Such second generation warm excess occurs around $\sim$ 
13\% of the early-type stars in nearby young stellar associations.
\end{abstract}


\section{Introduction}
In our Solar System, zodiacal dust grains are warm ($\gtrsim$150\,K) and 
found within $\sim$3\,AU of the Sun.  Slow but persistent collisions 
between asteroids complemented by material released from comets now 
replenish these particles.  Similar warm dust particles around other 
stars are also expected and would be manifested as excess mid-infrared 
emission.

The implication of ``warm'' excess stars for the terrestrial planet-building 
process has prompted many searches including several pointed observing
campaigns with {\it Spitzer} 
\citep{gor04,rie05,low05,bei05,bei06,bry06,hin06,gor06,smi06}.  
However, a lack of consensus of what constitutes a ``warm excess'' 
has resulted in ambiguity and some confusion in the field.  For example, 
Spitzer surveys with MIPS revealed a number of stars with excess 
emission in the 24\,$\micron$ band.  However, very few of these may 
turn out as genuine ``warm excess'' stars because the detected 
24\,$\micron$ emission is mostly the Wien tail of emission from cold 
(T $<$ 150\,K) dust grains \citep{rhe07a}.  

For black-body grains, 
$T_{dust}$ = T$_{*}$(R$_{*}$/(2R$_{dust})$)$^{1/2}$, where R$_{dust}$ 
is the distance of a grain from a star of radius R$_{*}$ and temperature 
T$_{*}$.  Due to the dependence of 
$T_{dust}$ on T$_{*}$ and R$_{*}$, the terrestrial planetary zone 
(TPZ) around high mass stars extends further out than that around 
low mass stars.  Therefore, R$_{dust}$ is not a good way to 
define the TPZ while dust equilibrium temperature is equally 
applicable to all main-sequence stars.  In our Solar system, T$_{dust}$ 
is 150K near the outer boundary of the asteroid belt ($\sim$3.5\,AU), 
and the zodiacal dust particles are sufficiently large 
($\sim$30\,$\micron$) that they do radiate like blackbodies.  To
specify a TPZ independent of the mass of the central star, we 
define the TPZ to be the region where T$_{dust} \gtrsim$ 150\,K.  
Then an A0 star has 25\,AU and an M0 star has 0.9\,AU as the outer 
boundary of their TPZ.  Because of the way it is defined, TPZ applies 
only to the location of grains that radiate like a blackbody.

According to the Spitzer surveys listed above, the presence of dust 
in the TPZ characterized by excess in the mid-IR is quite rare for 
stars $\gtrsim$10\,Myrs old.  For ages in the range of $8-30$\,Myr, a 
posited period of the terrestrial planet formation in our Solar System, 
only a few stars appear to possess warm dust according to our 
analysis (see $\S$ 5 and Table 1): $\eta$~Cha, 
a B8 member of 8 Myr old $\eta$ Cha cluster \citep{mam99}, $\eta$~Tel 
\& HD~172555, A0- and A7- type members of the 12 Myr old $\beta$ Pic 
moving group \citep{zuc04}, HD~3003, an A0 member of the 30 Myr old 
Tucana/Horologium moving group \citep{zsw01}, and HD~113766, an F3 
binary star (1.2$\arcsec$ separation, \citealt{dom94}), in the Lower 
Centaurus Crux (LCC) Association \citep{che05}.

In this paper, we present the A9 star EF Cha, another example of this 
rare group of stars with warm dust at the epoch of terrestrial planet 
formation.

\section{MSX Search for Mid-IR Excess Stars}

Hipparcos, 2MASS and Mid Course Experiment (MSX, \citealt{ega03}) sources 
were cross-correlated to identify main-sequence stars with excess 
emission at mid-IR wavelengths.  Out of $\sim$68,000 Hipparcos dwarfs with
$M_{V}$ $\geqq$ 6.0(\bv) - 2.0 (see \citealt{rhe07a} for an explanation 
of this $M_{V}$ constraint) in a search radius of 10$\arcsec$, $\sim$1000 
stars within 120 pc of Earth were identified with potential MSX 
counterparts.  

Spectral Energy Distributions (SED) were created for all $\sim$1,000 
MSX identified Hipparcos dwarfs.  Observed fluxes from Tycho-2 $B_{T}$ 
and $V_{T}$ and 2MASS $J$, $H$, and $K_{s}$, were fit to a stellar 
atmospheric model \citep{hau99} via a \(\chi\)$^{2}$ minimization method 
(see \citealt{rhe07a}, for detailed description of SED fitting).  From 
these SED fits, about 100 Hipparcos dwarfs were retained that 
showed apparent excess emission in the MSX 8\,$\micron$ band (that is, 
the ratio {[MSX flux - photosphere flux] / MSX flux uncertainty} must 
be $>$ 3.0).  Since a typical positional 3$\sigma$ uncertainty of MSX 
is $\sim$6\,$\arcsec$ \citep{cla05} and MSX surveyed the 
Galactic plane, a careful background check is required to eliminate 
contamination sources.  By over-plotting the 2MASS sources on the Digital 
Sky Survey (DSS) images, we eliminated more than half of the apparent 
excess stars that included any dubious object (i.e., extended objects, 
extremely red objects, etc.) within a 10$\arcsec$ radius from the star.  
Among the stars that passed this visual check, EF~Cha was selected for 
follow-up observations at the Gemini South Telescope.  Independent IRAS 
detections at 12 and 25\,$\micron$ made EF~Cha one of the best candidates 
for further investigation. 

\section{Ground-based Follow-up Observations \& MIPS Photometry}

An N-band image and a spectrum of EF~Cha were obtained using the
Thermal-Region Camera Spectrograph (T-ReCS) at the Gemini South Telescope
in March and July of 2006 (GS-2006A-Q-10), respectively.  Thanks to the
queue observing mode at Gemini Observatory, the data were obtained under
good seeing and photometric conditions. The standard ``beam switching''
mode was used in all observations in order to suppress sky emission and
radiation from the telescope.  Data were obtained chopping the secondary
at a frequency of 2.7 Hz and noddding the telescope every $\sim$30\,sec.  
Chopping and nodding were set to the same direction, parallel to the slit
for spectroscopy.

Standard data reduction procedures were carried out to reduce the 
image and the spectrum of EF~Cha at N-band.  Raw images were first 
sky-subtracted using the sky frame from each chop pair.  Bad pixels were 
replaced by the median of their neighboring pixels.  Aperture photometry 
was performed with a radius of 9 pixels (0.9\,$\arcsec$) and sky annuli 
of 14 to 20 pixels.  The spectrum of a standard star (HD 129078) was 
divided by a Planck function with the star's effective temperature 
(4500K) and this ratioed spectrum was then divided into the spectrum 
of EF~Cha to remove telluric and instrumental features.  The wavelength 
calibration was performed using atmospheric transition lines from an 
unchopped raw frame.  The 1-D spectrum was extracted by weighted 
averaging of 17 rows.

For the N-band imaging photometry, the on-source integration time of 
130 seconds produced S/N $>$ 30 with FWHM $\sim$0.54\,$\arcsec$. For 
the N-band spectrum, a 886 second on-source exposure resulted in S/N 
$>$ 20. A standard star, HIP 57092, was observed close in time and 
position to our target and was used for flux calibration of the N-band 
image of EF~Cha.  For spectroscopy, HD 129078, a KIII 2.5 star, was 
observed after EF~Cha at a similar airmass and served as a telluric 
standard.

While our paper was being reviewed, Spitzer Multiband Imaging Photometer 
for Spitzer (MIPS) archival images of EF~Cha at 24 and 70\,$\micron$ were 
released from the Gould's Belt Legacy program led by Lori Allen.  
EF~Cha was detected at MIPS 24\,$\micron$ band but not at MIPS 
70\,$\micron$ band.  We performed aperture photometry for EF~Cha at 
24\,$\micron$ on the post-BCD image produced by Spitzer Science Center MIPS 
pipeline.  We used aperture correction of 1.167 for the 24\,$\micron$ image 
given at the SSC MIPS website (http://ssc.spitzer.caltech.edu/mips/apercorr/) 
with aperture radius of 13\arcsec and sky inner and outer annuli of 20\arcsec 
and 32\arcsec, respectively.  For MIPS 70\,$\micron$ data, we estimated 
3~$\sigma$ upper limits to the non-detection on the mosaic image that we 
produced using MOPEX software on BCD images.  

\section{Results}
Table~2 lists the mid-IR measurements of EF~Cha from MSX, IRAS, MIPS, 
and Gemini T-ReCS observations.  The T-ReCS N-band image (FOV of
28.8$\arcsec$ $\times$ 21.6 $\arcsec$) confirmed that no other mid-IR 
source appears in the vicinity of EF~Cha and that the mid-IR excess 
detected by the space observatories (IRAS \& MSX) originates solely 
from EF~Cha.  A strong silicate emission feature in the N-band spectrum 
(Figure 1) indicates the presence of warm, small ($a$ $\lesssim$ 5$\micron$, 
see Figure 6 in \citealt{rhe06}) dust particles.  Amorphous silicate 
grains dominate the observed emission feature.  However, crystalline 
silicate structure, probably forsterite, appears as a small bump near 
11.3\,$\micron$ \citep{kes06}.  Polycyclic Aromatic Hydrocarbon 
(PAH) particles can also produce an emission feature at 11.3\,$\micron$.  
However, absence of other strong PAH emission features at 7.7 and 
8.6\,$\micron$ indicates that the weak 11.3\,$\micron$ feature does not 
arise from PAHs.  Furthermore, although PAH particles do appear in 
some very young stellar systems, they have not been detected around 
stars as old as 10 Myr.  In contrast, crystalline silicates 
such as olivine, forsterite, etc. are seen in a few such stellar systems 
\citep{son05,sch05,bei06,lis07}.  

The dust continuum excess of EF~Cha was fit with a single temperature 
blackbody curve at 240\,K by matching the flux density at 13\,$\micron$ 
and the MIPS 70\,$\micron$ upper limit (Figure 1).  The 3~$\sigma$ upper limit 
at MIPS 70\,$\micron$ band indicates that the dust temperature should not 
be colder than 240K.  Figure 1 shows that MIPS 24\,$\micron$ flux is 
$\sim$30\,mJy lower than IRAS 25\,$\micron$ flux.  Due to the small MIPS 
aperture size compared with IRAS, MIPS 24\,$\micron$ flux often comes out 
smaller when nearby contaminating sources are included in the large IRAS 
beam.  The ground-based T-ReCS (28.8\arcsec$\times$21.6\arcsec) image 
of EF~Cha at N-band, however, shows no contaminating source in the vicinity 
of EF~Cha.  Thus, the higher flux density at IRAS 25\,$\micron$ perhaps 
indicates the presence of a significant silicate emission feature near 
18\,$\micron$ included in the wide passband of the IRAS 25\,$\micron$ filter 
(18.5 $-$ 29.8\,$\micron$).  A recent Spitzer IRS observation of another 
warm excess star, BD+20 307 \citep{son05}, shows a similar discrepancy 
between the IRAS 25\,$\micron$ flux and MIPS 24\,$\micron$ flux in the presence of a 
significant silicate emission feature at $\sim$18\,$\micron$ (Weinberger et 
al.\, 2007 in preparation), consistent with our interpretation.  
MIPS 24\,$\micron$ flux is slightly above our 240\,K dust continuum fit.  
The wide red wing of an 18\,$\micron$ silicate emission feature could 
contribute to a slight increase in MIPS 24\,$\micron$ flux.

\section{Discussion}
\subsection{Debris Disk Characteristics of EF~Cha}
EF~Cha was detected in the ROSAT X-ray All Sky Survey with $L_{x}/L_{bol} 
= 10^{-4.68}$ which suggests a very young age for an A9 star (see Figure 
4 in \citealt{zuc04}).  On the basis of this X-ray measurement, Hipparcos 
distance (106 pc), location in the sky (RA = 12$^\circ$07$^\prime$, Dec= 
-79$^\circ$), and proper motion (pmRA = -40.2$\pm$1.2 \& pmDE = 
-8.4$\pm$1.3 in mas/yr), EF~Cha is believed to be a member of the 
``Cha-Near'' moving group (avg. RA = 12$^\circ$00$^\prime$ \& avg. DEC = 
-79$^\circ$, avg. pmRA = -41.13$\pm$1.3 \& avg. pmDEC = -3.32$\pm$0.86 
in mas/yr, \citealt{zuc04,zuc07}), which is $\sim$10 Myr old and typically 
$\sim$90pc from Earth.

Large blackbody grains in thermal equilibrium at 240\,K, would be located 
$\sim$4.3\,AU from EF Cha while small grains, especially those 
responsible for the silicate emission features in our N-band spectrum, 
radiate less efficiently and could be located at $>$4.3\,AU.  Recent Spitzer 
MIPS observations confirmed that all aforementioned ($\S$ 1) warm excess 
stars do not have a cold dust population, indicating few large grains at 
large distances \citep{che06,smi06}.  Lack of cold large grains, in turn, 
suggests local origin of the small grains seen in these warm excess stars.  
Without cold excess from Spitzer MIPS 70\,$\micron$ data, small grains in 
EF~Cha should originate in the TPZ, probably by the breakup of large grains 
in the TPZ, rather than inward migration from an outer disk.  Even in the 
unlikely event that silicate emission comes from small grains in an 
outer disk that were blown away by radiation pressure as in Vega \citep{su05}, 
the dominant carrier of 240\,K continuum emission would still be large grains 
(Aigen Li 2007, private communication).

The fraction of the stellar luminosity reradiated by dust, $\tau$, is 
$\sim 10^{-3}$, which was obtained by dividing the infrared excess 
between 7\,$\micron$ and 60\,$\micron$ by the bolometric stellar luminosity.  
This $\tau$ is $\sim$10,000 times larger than that of the current Sun's 
zodiacal cloud ($\sim 10^{-7}$) but appears to be moderate for known 
debris disk systems at similar ages (see Figure 4 in \citealt{rhe07a}).  
\citet{rhe07a} show that the ratio of dust mass to $\tau$ of a debris 
system is proportional to the inverse-square of dust particle semimajor 
axis for semimajor axes between $\sim$9\,AU and $\sim$100\,AU.  For systems 
with dust radius $<$\,9\,AU, this relationship overestimates the dust mass.  
Instead, we calculate the mass of a debris ring around EF~Cha using 
\begin{equation}
M_{dust} \geq \frac{16}{3} \pi \frac{L_{IR}}{L_{*}} \rho_{s} R_{dust}^{2} <a> 
\end{equation}
(eqn. 4 in \citealt{che01}), where $\rho_{s}$ is the density of an 
individual grain, $L_{IR}$ is the dust luminosity, and $<a>$ is the 
average grain radius.  Because \citet{che01} analyzed $\zeta$~Lep, a 
star of similar spectral type to EF~Cha, we adopt their model for grain 
size distribution.  Assuming R$_{dust}$ = 4.3\,AU, $\rho_{s}$ = 2.5 
g~$cm^{-3}$, and $<a>$ = 3\,$\micron$, the dust mass is $4.8 \times 
10^{22}$ g ($\sim$10$^{-5}$ $M_{\earth}$).  Grains with $a$ $=$ 
3\,$\micron$ will radiate approximately as blackbodies at wavelengths 
shorter than $\sim$2$\pi$$a$ ($\sim$20\,$\micron$).  As may be seen 
from Figure 1, most of the excess IR emission at EF~Cha appears at 
wavelengths $\lesssim$20\,$\micron$.

For blackbody grains at $\sim$4.3\,AU with, for example, radius 
$a$ $=$ 3\,$\micron$, the Poynting Robertson (P-R) drag time scale
\footnote{In the calculation of P-R timescale, 9.7 $L_{\sun}$ was used 
for the stellar luminosity of EF~Cha with a bolometric correction of 
-0.102 \citep{cox00}.  Absolute visual 
magnitude of EF~Cha is 2.35 based on the Hipparcos distance of 106 pc.  } 
is only $8\times10^{3}$ years, much less than 
10 Myrs.  Yet smaller grains with $a <$ 1.3\,$\micron$ would be easily 
blown away by radiation pressure on a much shorter time-scale.  
Successive collisions among grains can effectively remove dust 
particles by grinding down large bodies into smaller grains, which then 
can be blown out.  The characteristic collision time (orbital 
period/$\tau$) of dust grains at 4.3\,AU from this A9 star is $\sim 10^{4}$ 
years.  Both PR time and collision time were derived assuming no gas was 
present in the disk.  While gas has not been actively searched for in 
EF~Cha, few debris disk systems at $\gtrsim$ 10 Myr show presence of gas, 
indicating early dispersal of gas \citep{pas06}.  A possibility of an 
optically thin gas disk surviving around a $\sim$10 Myr system 
was investigated by \citet{tak01}.  However their model of gas disk 
is pertinent to cool dust at large distances ($\gtrsim$ 120 AU), but 
not to warm dust close to the central star as in EF~Cha.  In addition, 
a recent study of OB associations shows that the lifetime of a primordial 
inner disk is $\lesssim$ 3\,Myr for Herbig Ae/Be stars \citep{her05}.  
Based on the very short time scales of dust grain removal, essentially 
all grains responsible for significant excess emission at EF~Cha in the 
mid-IR are, therefore, likely to be second generation, not a remnant of 
primordial dust.  

\subsection{Debris Disk Systems in the TPZ during the Epoch of Planet 
Formation} 

The presence of hot dust has been recognized around other $\sim$10\,Myr
stars, for example, TW~Hya, HD~98800, \& Hen~3-600 in the TW Hydrae
Association.  Interesting characteristics of these systems are their large
$\tau$ ($\gtrsim10^{-2}$) and late-K and M spectral types.  TW~Hya and
Hen~3-600 show flat IR SED up to 160\,$\micron$ consistent with active
accretion in their disks \citep{low05}.  Combined with the presence of
substantial gas emission lines from TW~Hya, the observed infrared excess
emission, at least for these two stars, appears to arise from gaseous dusty
disks left over from the protostellar environment.  On the other hand, a
lack of gas emission (\citealt{den05}) and the quadruple nature of the HD
98800 system has invoked a flared debris disk as an alternative
explanation for its large infrared excess emission \citep{fur07}.  Many
young stars come in multiple systems.  However, they hardly display such 
a high $\tau$ as that of HD~98800.  Thus, the dust disk around HD~98800 
might be an unusual transient pheonomeon such that it still contains a 
dust population composed of a mixture of promordial grains and replenished 
debris.  Some stars display a mixture of warm and cold grains where the 
overall infrared excess emissions is dominated by the cold dust.  
Table~1 summarizes the currently known disk systems with warm dust 
regardless of spectral type and the presence of cold excess or remnant 
primordial dust at $\sim 8-30$\,Myr.

What separates EF~Cha from the stars described in the previous paragraph 
is that most of the infrared excess emission, if not all, arises from 
warm dust in the TPZ, and as described in $\S$\,5.1, these grains are, 
clearly, not a remnant of the protostellar disk.  Recent Spitzer MIPS 
observation shows null detection of EF Cha at 70\,$\micron$ band, leaving  
the presence of substantial cold dust unlikely (See Figure~1).  This result 
is consistent with the recent Spitzer observations of other similar Table~1 
early-type warm excess systems ($\eta$~Tel, HD~3003, HD~172555 and HD~113766) 
in which cold dust from a region analogous to the Sun's Kuiper belt 
objects is missing \citep{che06,smi06}.  In the following discussion, 
we characterize warm excess stars as those with warm dust in the TPZ 
only and without cold excess (i.e., we exclude stars like $\beta$ Pictoris).

The fact that all currently known warm excess stars at ages between 
$8-30$ Myr belong to nearby stellar moving groups offers 
an excellent opportunity to address how frequently warm excess emission 
appears among young stars in the solar neighborhood.  \citet{zuc04} 
list suggested members of stellar moving groups and 
clusters (i.e., $\eta$~Cha cluster, TW Hydra Association, $\beta$~Pictoris 
Moving Group, Cha-Near Moving Group, and Tucana/Horologium Association) 
at ages $8-30$ Myr, within 100 pc of Earth.  Currently Spitzer MIPS 
archive data are available for all 18 members of the $\eta$ Cha cluster, 
all 24 members of TWA, all 52 members of Tucana/Horologium, 25 out of 
27 $\beta$ Pic Moving Group, and 9 out of 19 members of Cha-Near Moving 
Group.  Multiple systems were counted as one object unless resolved by 
Spitzer.  For example, in the $\beta$ Pictoris Moving Group, HD~155555A, 
HD~155555B \& HD~155555C were counted as a single object; however, 
HIP~10679 \& HIP~10680 were counted as two objects.  

Table 1 shows that the characteristics of dust grains depend on the 
spectral type of the central star.  All six warm debris disks considered 
in this paper harbor early-type central stars (earlier than F3).  
However, late-type stars in Table 1 (for example, PDS 66, a 
K1V star from LCC) sometimes show characteristics of T~Tauri-like 
disk excess (i.g., $\tau >10^{-2}$, flat IR SED, etc.) even at 
$\gtrsim$10 Myr \citep{sil06}.  Such apparent spectral dependency 
perhaps arises from the relatively young ages ($\sim$10\,Myr) of 
these systems in which late-type stars still possess grains mixed with 
primordial dust due to a longer dust removal time scale \citep{her05}. 
In the above-mentioned five nearby stellar moving groups, 38 out of 
129\footnote{128 stars with Spitzer data plus EF~Cha.  EF~Cha, which appeared 
in \citet{mam00}, was inadvertently omitted from the suggested members of 
Cha-Near Moving Group by \citet{zuc04}, but its membership is included in 
Zuckerman et al. (2007, in prep.).  Since the warm excess in EF~Cha was 
found by MSX \& IRAS surveys less-sensitive than {\it Spitzer} MIPS, the 
addition of EF~Cha to the homogenious pool of MIPS surveyed stars 
does not increase the warm excess occurrence rate inappropriately because 
MIPS would have easily detected it.  In the near future, even if new 
members are added to these nearby stellar associations, we believe 
the overall occurrence rate would remain similar to the current estimate 
because most A- through early-M type members of these stellar associations, 
excepting perhaps the Cha-Near Moving Group, are already well established 
through extensive searches and almost all the members are already 
surveyed by {\it Spitzer} MIPS.} stars with Spitzer MIPS measurements have spectral types earlier 
than G0.  Therefore, we find $\sim$13\% (5/38) occurrence rate for 
the warm excess phenomenon among the stars with spectral types earlier 
than G0 in the nearby stellar groups at $8-30$ Myr.  (Beta Pic is the only
early-type star among the remaing 33 which has both warm and cold dust.)  
For LCC, at least one (HD~113766) out of 20 early-type members is a warm 
excess star giving 5\% frequency (see Table 1 \& 2 in \citealt{che05}).  
This rate can reach a maximum of 30\% when we take into account five 
early-type LCC members that show excess emission at 24\,$\micron$ but 
have only upper-limit measurements at MIPS 70\,$\micron$ band.  G0 type 
was chosen to separate the two apparently different populations because 
no G-type star except T~Cha appears in Table 1.  Furthermore, the spectral 
type of T~Cha is not well established (G2-G8) and it may be a K-type star 
like other K/M stars in Table 1.  Rhee et al.\ (2007, in preparation) 
analyze all spectral types in the young nearby moving groups and conclude 
that the warm excess phenomenon with $\tau \gtrsim 10^{-4}$ occurs for 
between 4 and 7\%; this uncertainty arises because some stars have only 
upper limits to their 70\,$\micron$ fluxes.

\acknowledgments

We thank Aigen Li for helpful advice and the referee for constructive 
comments that improved the paper.  
This research was supported by NASA grant NAG5-13067 to Gemini 
Observatory, a NASA grant to UCLA, and Spitzer GO program \#3600.  
Based on observations obtained at the Gemini Observatory, which is 
operated by the Association of Universities for Research in Astronomy, 
Inc., under a cooperative agreement with the NSF on behalf of the 
Gemini partnership: the National Science Foundation (United States), 
the Particle Physics and Astronomy Research Council (United Kingdom), 
the National Research Council (Canada), CONICYT (Chile), the Australian 
Research Council (Australia), CNPq (Brazil) and CONICET (Argentina).  
This research has made use of the VizieR catalogue 
access tool, CDS, Strasbourg, France and of data products from the Two 
Micron All Sky Survey (The latter is a joint project of the University 
of Massachusetts and the Infrared Processing and Analysis 
Center/California Institute of Technology, funded by the National 
Aeronautics and Space Administration and the National Science Foundation).


\clearpage

\begin{deluxetable}{ccccccccccccccccc}
\tabletypesize{\tiny}
\setlength{\tabcolsep}{0.025in}
\tablecaption{Debris systems in the terrestrial planetary zone in the epoch of terrestrial planet formation}
\tablewidth{0pt}
\tablehead{
\colhead{}                    &\colhead{}           &
\colhead{V}                   &\colhead{D}          &\colhead{$R_{star}$\tablenotemark{a}}   &
\colhead{$T_{star}$} 	      &\colhead{$T_{dust}$} &
\colhead{$\tau$}     	      &\colhead{age}        &\colhead{Warm excess}             &
\colhead{Disk}                &\colhead{}           &\colhead{}            \\
\colhead{Star}       	      &\colhead{Sp. Type}   &
\colhead{(mag)}      	      &\colhead{(pc)}       &\colhead{($R_\odot$)}  &
\colhead{(K)}        	      &\colhead{(K)}        &
\colhead{($\times10^{-4}$)}   &\colhead{(Myr)}      &\colhead{Only}  &
\colhead{Characteristics}     &\colhead{Membership} &\colhead{References}  \\
\colhead{(1)}        	      &\colhead{(2)}        &
\colhead{(3)}        	      &\colhead{(4)}        &\colhead{(5)}          &
\colhead{(6)}        	      &\colhead{(7)}        &
\colhead{(9)}        	      &\colhead{(10)}       &\colhead{(11)}         &
\colhead{(12)}                &\colhead{(13)}       &\colhead{(14)}        \\
}
\startdata
$\eta$ Cha & B8\tablenotemark{b} & 5.5 & 97 & 2.37 & 11000    & 320  & 1.5   & 8    & Yes  & Debris & $\eta$ Cha  & 1 \\
EF Cha      	  & A9   & 7.5  & 106       & 1.92 & 7400     & 240  & 10    & 10   & Yes  & Debris & Cha-Near    & 2\\
HD 113766  	  & F3V  & 7.5  & 131       & ...\tablenotemark{c} & 7000     & 350  & 150   & 10\tablenotemark{d}   & Yes  & Debris & LCC  & 3,5,6\\
HD 172555   	  & A7V  & 4.8  & 29        & 1.52 & 8000     & 320  & 8.1   & 12   & Yes  & Debris & $\beta$ Pic & 3,4,6\\
$\eta$ Tel  	  & A0V  & 5.0  & 48        & 1.61 & 9600     & 150  & 2.1   & 12   & Yes  & Debris & $\beta$ Pic & 3,6 \\   
HD 3003           & A0V  & 5.1  & 46        & 1.59 & 9600     & 200  & 0.92  & 30   & Yes  & Debris & Tucana/Horologium & 7,8\\
\\
$\beta$ Pic       & A5V  & 3.9  & 19        & 1.37 & 8600     & 110\tablenotemark{e}  & 26    & 12   & No   & Debris & $\beta$ Pic & 4 \\
HD 98800    	  & K5Ve & 9.1  & 47        & ...\tablenotemark{c} & 4200     & 160  & 1100  & 8    & No   & Primordial/Debris  &  TWA   	    & 2,7\\
TW Hya      	  & K8Ve & 11.1 & 56        & 1.11 & 4000     & 150? & $>$2200  & 8    & No   & Primordial/Debris  &  TWA        & 2,7\\
Hen 3-600 & M3 & 12.1 & 42\tablenotemark{f} & ...\tablenotemark{c} & 3200     & 250  & 1000  & 8    & No   & Primordial/Debris  &  TWA   	    & 2,7\\
EP Cha      	  & K5.5\tablenotemark{b}   & 11.2 & 97  & 1.39 & 4200\tablenotemark{b}     & ...\tablenotemark{g}  & $>$1300\tablenotemark{g}  & 8    & No   & Primordial/Debris  & $\eta$ Cha  & 10,11\\
ECHA J0843.3-7905 & M3.25\tablenotemark{b}  & 14.0 & 97  & 0.87 & 3400\tablenotemark{b}     & ...\tablenotemark{g}  & $>$2000\tablenotemark{g}  & 8    & No   & Primordial/Debris  & $\eta$ Cha  & 10,11\\
EK Cha        	  & M4\tablenotemark{b}     & 15.2 & 97  & 0.79 & 3300\tablenotemark{b}     & ...\tablenotemark{g}  & $>$590\tablenotemark{g}   & 8    & No   & Primordial/Debris  & $\eta$ Cha  & 10,11\\   
EN Cha      	  & M4.5\tablenotemark{b}   & 15.0 & 97  & ...\tablenotemark{c} & 3200\tablenotemark{b}     & ...\tablenotemark{g}  & $>$480\tablenotemark{g}   & 8    & No   & Primordial/Debris  & $\eta$ Cha  & 10,11\\
ECHA J0841.5-7853 & M4.75\tablenotemark{b}  & 14.4 & 97  & 0.51 & 3200\tablenotemark{b}     & ...\tablenotemark{g}  & $>$480\tablenotemark{g}   & 8    & No   & Primordial/Debris  & $\eta$ Cha  & 10,11\\
ECHA J0844.2-7833 & M5.75\tablenotemark{b}  & 18.4 & 97  & 0.40 & 3000\tablenotemark{b}     & ...\tablenotemark{g}  & $>$400\tablenotemark{g}   & 8    & No   & Primordial/Debris  & $\eta$ Cha  & 10,11\\
T Cha             & G2-G8                   & 11.9 & 66  & ...\tablenotemark{h}  & ...\tablenotemark{h}  & ...\tablenotemark{h}  & ...\tablenotemark{h}   & 10 & No & Primordial/Debris  & Cha-Near & 7,12 \\
PDS 66            & K1Ve & 10.3 &$\sim$85   & 1.35\tablenotemark{i} & 4400  & ...\tablenotemark{g}  & $>$2200\tablenotemark{g}& 10\tablenotemark{d}   & No   & Primordial/Debris  & LCC   & 6,9  \\
\enddata
\tablenotetext{a}{Estimated using a total integrated flux for a given distance (col. 4) assuming that each star is a single object.}
\tablenotetext{b}{From \citet{luh04}}
\tablenotetext{c}{Multiple systems: HD 113766 (binary), HD 98800 (quadruple), Hen 3-600 (triple), \& EN Cha (binary) from \citet{zuc04}.}
\tablenotetext{d}{\citet{mam02} estimate the age of LCC at $\sim$16 Myr.  However, \citet{son07} report a younger age of $\sim$10 Myr.}
\tablenotetext{e}{Based on the single temperature blackbody fit to the dominant cold excess.  
However, additional warm excess emission exists above the 110K model fit, indicating the presence 
of warm disk in terrestrial planetary zone.}
\tablenotetext{f}{Estimated photometric distance from \citet{zuc04}}
\tablenotetext{g}{This lower limit $\tau$ is estimated from our two-temperature dust component fit to the infrared excess emission.  
No dust temperature is given because its IR excess emission cannot be fit 
with a single temperature blackbody.  However, a significant excess emission at mid-IR 
wavelengths exists which is not the Wien tail of emission from cold dust grains.}
\tablenotetext{h}{No SED fitting was attempted to estimate stellar radius \& temperature as well as dust parameters.}
\tablenotetext{i}{Based on the distance of 85 pc \citep{sil06} from Earth}
\tablecomments{$R_{star}$ and $T_{star}$ were obtained by fitting the observed 
optical and near-IR measurements with NextGen stellar atmosphere model \citep{hau99}.  
$T_{dust}$ and $\tau$ were estimated by fitting blackbody curves to the infrared excess emission. \\
Refereces: 
1. \citet{mam99}
2. \citet{zuc07}
3. \citet{sch05}
4. \citet{zuc01}
5. \citet{che05}
6. \citet{che06}
7. \citet{zuc04} 
8. \citet{smi06}
9. \citet{sil06}
10. \citet{mam02}
11. \citet{meg05}
12. \citet{kes06}}
\end{deluxetable}

\clearpage

\begin{deluxetable}{cccccccc}
\tablecolumns{8}
\tablecaption{Mid-IR measurements of EF~Cha \label{tbl-1}}
\tabletypesize{\small}
\tablewidth{0pc}
\tablehead{
\colhead{}               &\colhead{Central}          &
\colhead{}               &\colhead{Adopted Photospheric} &
\colhead{Excess}         &\colhead{}   \\
\colhead{}               &\colhead{Wavelength}          &
\colhead{Flux Density}   &\colhead{Flux Density} &
\colhead{Flux Density}   &\colhead{}   \\
\colhead{Band}           &\colhead{($\micron$)}         &
\colhead{(mJy)}          &\colhead{(mJy)}               &
\colhead{(mJy)}               &\colhead{Instrument} 
}
\startdata
8$\micron$         & 8.28  & 167 $\pm$9    & 119 & 48 & MSX            \\
N                  & 10.4  & 164 $\pm$14   &  84 & 80 & Gemini/TReCS   \\
N                  & 7.7-12.97  & --       &  -- & -- & TReCS, $\lambda$/$\Delta\lambda \approx$ 100 spectrum   \\
12$\micron$        & 11.5  & 152 $\pm$32   &  58 & 94 & IRAS           \\
24$\micron$        & 24.0  &  80 $\pm$4    &  14 & 66 & MIPS           \\
25$\micron$        & 23.7  & 110 $\pm$21   &  14 & 96 & IRAS           \\
70$\micron$        & 70.0  &  $<$ 22.4\tablenotemark{a} & 1.7 & $<$ 20.7  & MIPS    \\
\enddata
\tablenotetext{a}{3~$\sigma$ upper limit to the non-detection.}
\tablecomments{Both MSX and IRAS flux densities were color-corrected 
using the method described in \citet{rhe07a}.}
\end{deluxetable}

\clearpage

\begin{figure}
\plotone{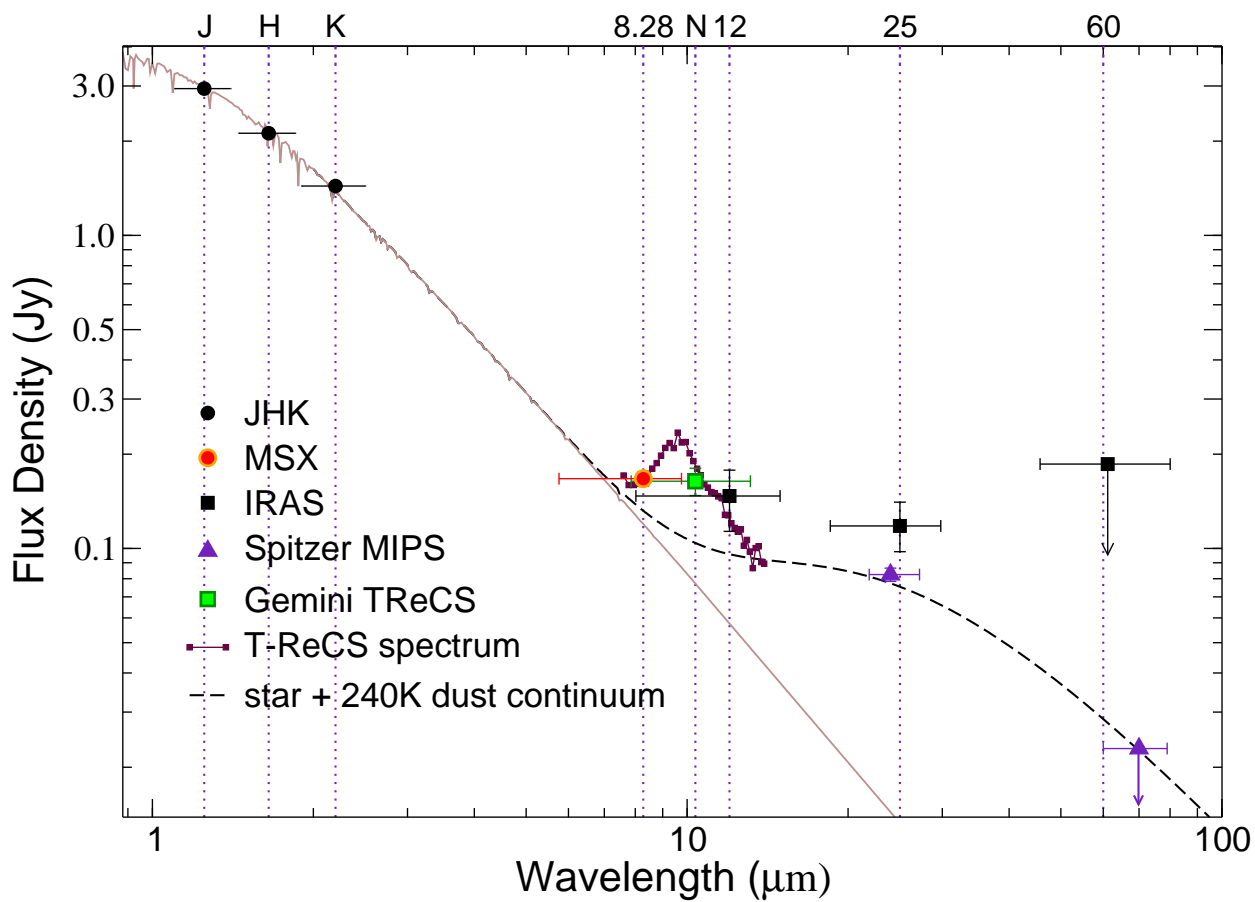}
\caption{Spectral energy distribution of EF~Cha created 
by fitting existing optical and near-IR measurements with a 7400K stellar 
atmosphere model.  After subtracting the stellar photosphere as well as 
the silicate emission feature near 10\,$\micron$, the infrared excess 
continuum emission was fit with a single temperature blackbody model of 
240\,K.  For each measurement, the horizontal bars indicate the passband 
of the filter used and the vertical bars depict the flux uncertainties.  
Flux density values for the plotted mid-IR points are given in Table 2.}
\label{SED}
\end{figure}


\begin{thebibliography}{}
\bibitem[Acke \& van den Ancker(2004)]{ack04} Acke, B., \& van den 
  Ancker, M.~E.\ 2004, \aap, 426, 151 
\bibitem[Beichman et al.(2005)]{bei05} Beichman, C.~A., et 
  al.\ 2005, \apj, 626, 1061 
\bibitem[Beichman et al.(2006)]{bei06} Beichman, C.~A., et 
  al.\ 2006, \apj, 639, 1166 
\bibitem[Bryden et al.(2006)]{bry06} Bryden, G. et al. 2006, ApJ, 
  636, 1098
\bibitem[Chen \& Jura(2001)]{che01} Chen, C.~H., \& Jura, M.\ 
  2001, \apjl, 560, L171 
\bibitem[Chen et al.(2005)]{che05} Chen, C.~H., Jura, M., Gordon, 
  K.~D., \& Blaylock, M.\ 2005, \apj, 623, 493 
\bibitem[Chen et al.(2006)]{che06} Chen, C.~H., et al.\ 2006, \apjs, 
  166, 351 
\bibitem[Clarke et al.(2005)]{cla05} Clarke, A.~J., 
   Oudmaijer, R.~D., \& Lumsden, S.~L.\ 2005, \mnras, 363, 1111 
\bibitem[Cox(2000)]{cox00} Cox, A.~N.\ 2000, Allen's 
   Astrophysical Quantities, 4th ed.~Edited by Arthur N.~Cox.~Publisher: 
   New York: AIP Press; Springer, 2000.~ISBN: 0387987460  
\bibitem[Dent et al.(2005)]{den05} Dent, W.~R.~F., Greaves, 
   J.~S., \& Coulson, I.~M.\ 2005, \mnras, 359, 663 
\bibitem[Dommanget \& Nys(1994)]{dom94} Dommanget, J., \& Nys, O.\ 
   1994, Communications de l'Observatoire Royal de Belgique, 115, 1 
\bibitem[Egan et al.(2003)]{ega03} Egan, M.~P., et al.\ 2003, 
   VizieR Online Data Catalog, 5114, 0 
\bibitem[Furland et al.(2007)]{fur07} Furland et al. 2007, ApJ in 
   press, Astro-ph/07050380.
\bibitem[Gorlova et al.(2004)]{gor04} Gorlova N., et al. 2004, ApJS, 
  154, 448
\bibitem[Gorlova et al.(2006)]{gor06} Gorlova, N., Rieke, G.~H., 
  Muzerolle, J., Stauffer, J.~R., Siegler, N., Young, E.~T., \& 
  Stansberry, J.~H.\ 2006, \apj, 649, 1028 
\bibitem[Hauschildt et al.(1999)]{hau99} Hauschildt, P.~H., 
   Allard, F., \& Baron, E.\ 1999, \apj, 512, 377 
\bibitem[Hern{\'a}ndez et al.(2005)]{her05} Hern{\'a}ndez, J., Calvet, 
   N., Hartmann, L., Brice{\~n}o, C., Sicilia-Aguilar, A., \& Berlind, 
   P.\ 2005, \aj, 129, 856 
\bibitem[Hines et al.(2006)]{hin06} Hines, D. C. et al. 2006, \apj, 
  638, 1070
\bibitem[Kessler-Silacci et al.(2006)]{kes06} Kessler-Silacci, J., 
   et al.\ 2006, \apj, 639, 275 
\bibitem[Lisse et al.(2006)]{lis07} Lisse, C. M., Beichman, C. A., 
   Bryden, G., \& Wyatt, M. C. 2007, \apj, 658, 584
\bibitem[Low et al.(2005)]{low05} Low, F., Smith, P. S., Werner, M., 
  Chen, C., Krause, V., Jura, M., \& Hines, D. C. 2005, ApJ 631, 1170 
\bibitem[Luhman \& Steeghs(2004)]{luh04} Luhman, K.~L., \& 
  Steeghs, D.\ 2004, \apj, 609, 917 
\bibitem[Mamajek et al.(1999)]{mam99} Mamajek, E. M, Lawson, W. A., 
   \& Feigelson, E. D. 1999, \apj, 516, L77
\bibitem[Mamajek et al.(2000)]{mam00} Mamajek, E.~E., Lawson, W.~A., 
   \& Feigelson, E.~D.\ 2000, \apj, 544, 356 
\bibitem[Mamajek et al.(2002)]{mam02} Mamajek, E.~E., Meyer, 
   M.~R., \& Liebert, J.\ 2002, \aj, 124, 1670 
\bibitem[Megeath et al.(2005)]{meg05} Megeath, S.~T., 
   Hartmann, L., Luhman, K.~L., \& Fazio, G.~G.\ 2005, \apjl, 634, L113 
\bibitem[Pascucci et al.(2006)]{pas06} Pascucci, I., et al.\ 
   2006, \apj, 651, 1177 
\bibitem[Rhee \& Larkin(2006)]{rhe06} Rhee, J.~H., \& Larkin, J.~E.\ 
   2006, \apj, 640, 625 
\bibitem[Rhee et al.(2007)]{rhe07a} Rhee, J. H., Song, I., Zuckerman, B.,
   \& McElwain, M. 2007, \apj, 660, 1556
\bibitem[Rieke et al.(2005)]{rie05} Rieke, G.~H., et al.\ 
   2005, \apj, 620, 1010 
\bibitem[Schuetz et al.(2005)]{sch05} Schuetz, O., Meeus, G., \& Sterzik, 
   M. F. 2005, \aap, 431, 175
\bibitem[Silverstone et al.(2006)]{sil06} Silverstone, M. D. et al. 2006, 
   \apj, 639, 1138 
\bibitem[Smith et al.(2006)]{smi06} Smith, P.~S., Hines, 
D.~C., Low, F.~J., Gehrz, R.~D., Polomski, E.~F., \& Woodward, C.~E.\ 2006, 
\apjl, 644, L125 
\bibitem[Song et al.(2005)]{son05} Song, I., Zuckerman, B., Weinberger, 
   A. J., Becklin, E. E. 2005, Nature, 436, 363
\bibitem[Song et al.(2007)]{son07} Song, I., Zuckerman, B., Bessell, M. 
   2007, submitted to ApJ.
\bibitem[Su et al.(2005)]{su05} Su, K.~Y.~L., et al.\ 2005, \apj, 628, 487 
\bibitem[Takeuchi \& Artymowicz(2001)]{tak01} Takeuchi, T., \& 
  Artymowicz, P.\ 2001, \apj, 557, 990 
\bibitem[Wyatt et al.(2007)]{wya07} Wyatt, M.~C., Smith, R., Greaves, 
   J.~S., Beichman, C.~A., Bryden, G., \& Lisse, C.~M.\ 2007, \apj, 
   658, 569
\bibitem[Zuckerman et al.(2001)]{zuc01} Zuckerman, B., Song, 
I., Bessell, M.~S., \& Webb, R.~A.\ 2001, \apjl, 562, L87 
\bibitem[Zuckerman et al.(2001)]{zsw01} Zuckerman, B., Song, 
I., \& Webb, R.~A.\ 2001, \apj, 559, 388 
\bibitem[Zuckerman \& Song(2004)]{zuc04} Zuckerman, B., \& 
Song, I.\ 2004, \araa, 42, 685 
\bibitem[Zuckerman et al.(2007)]{zuc07} Zuckerman, B., Song, I., 
   Weinberger, A., Bessell, M. 2007 ApJ Letter in preparation. 
\end{thebibliography}
\end{document}